\documentclass[prc,preprints, showpacs,amsmath,amssymb,superscriptaddress,floatfix,nofootinbib,10pt]{revtex4}
\usepackage{mathrsfs,bm,bbm}
\usepackage{longtable,lscape}
\usepackage{slashed,slashbox}
\usepackage{array}
\usepackage{txfonts}
\usepackage{indentfirst}
\usepackage{graphicx,booktabs}
\usepackage{multirow}
\usepackage{overpic}
\usepackage{color}

\begin{document}
\title{Covariant chiral  nucleon-nucleon contact Lagrangian up to order $\mathcal{O}(q^4)$}

\author{Yang Xiao}
\affiliation{School of Physics and
Nuclear Energy Engineering \& International Research Center for Nuclei and Particles in the Cosmos \&
Beijing Key Laboratory of Advanced Nuclear Materials and Physics,  Beihang University, Beijing 100191, China}

\author{Li-Sheng Geng}
\email[E-mail: ]{lisheng.geng@buaa.edu.cn} \affiliation{School of
Physics and Nuclear Energy Engineering \& International Research
Center for Nuclei and Particles in the Cosmos \& Beijing Key
Laboratory of Advanced Nuclear Materials and Physics,  Beihang
University, Beijing 100191, China}

\author{Xiu-Lei Ren}
\affiliation{Ruhr-Universit\"{a}t Bochum, Fakult\"{a}t f\"{u}r Physik und Astronomie, Institut f\"{u}r Theoretische Physik II, D-44780 Bochum, Germany}

\begin{abstract}
We adopt a covariant version of the naive dimensional analysis and construct the chiral two-nucleon contact Lagrangian constrained by Lorentz,
parity, charge conjugation, hermitian conjugation, and chiral symmetries. We show that at $\mathcal{O}(q^0)$, $\mathcal{O}(q^2)$, $\mathcal{O}(q^4)$, where $q$ denotes a generic small quantity,
there are 4, 13, and 23 terms, respectively.  We find that by performing $1/m_N$ expansions, the covariant Lagrangian reduces to the conventional non-relativistic one, which includes 2, 7, and 15 terms at each corresponding order.

\end{abstract}


\maketitle
\section{INTRODUCTION}

Chiral perturbation theory ($\chi$PT), first proposed by Weinberg~\cite{Weinberg:1968de,Weinberg:1978kz}, has
turned out to be a very successful tool in the study of the non-perturbative strong interaction physics. As a low-energy effective field theory of QCD, the theory of the strong interaction,
it is consistent with all the relevant symmetries, particularly, the chiral symmetry and its breaking pattern. The interactions between the Nambu-Goldstone bosons themselves and with
a heavier hadron can be computed in a systematic way order by order as an expansion of $q/\Lambda_\chi$, where $q$ denotes a generic small quantity and $\Lambda_\chi$ is the chiral symmetry breaking scale.
The unknown short-range physics is encoded in the so-called low-energy constants (LECs), which in principle can be obtained by matching with QCD, but in practice are often determined
by either fitting to experimental data or lattice QCD simulations. The predictive power of $\chi$PT relies on the fact that for specific observables only a selected set of LECs contribute and with the same LECs it relates different observables.

$\chi$PT has been successfully applied  in the pure mesonic sector~\cite{Gasser:1983yg,Gasser:1984gg}. However, when expanded to the one baryon sector, one encounters a tricky problem since the relatively large baryon mass cannot be counted as a small expansion parameter,
which does not vanish in the chiral limit and  breaks the power counting first introduced in the mesonic sector.  In the past two decades, several modified power counting schemes have been developed to overcome this problem.
At the very beginning, a non-relativistic approach, $i.e.$, the heavy baryon (HB) chiral perturbation theory~\cite{Jenkins:1990jv,Bernard:1995dp}, inspired by the heavy quark effective field theory~\cite{Georgi:1990um},
was introduced. More recently  two relativistic approaches, i.e., the infrared (IR)~\cite{Becher:1999he} and the extended-on-mass-shell (EOMS) scheme~\cite{Gegelia:1999gf,Fuchs:2003qc} were
developed to overcome some of the drawbacks of the non-relativistic approach. The latter seems to be successful both
formerly and empirically~\cite{Geng:2008mf,Geng:2009ik,Lensky:2009uv,Geng:2010vw,Alarcon:2012kn,Chen:2012nx,Ren:2012aj,Altenbuchinger:2013vwa,Ren:2013oaa,Geng:2014efa,Ledwig:2014rfa,Blin:2014rpa,Ren:2014vea,Yao:2015qia,Sun:2016wzh,Ling:2017jyz,Xu:2017tsr,Xiao:2018rvd,Yao:2018ifh,Blin:2018pmj,Liu:2018euh} .~\footnote{For a concise review, see, e.g., Ref.~\cite{Geng:2013xn}.}

In the 1990's, Weinberg proposed that one can construct the nucleon-nucleon interaction using heavy baryon chiral perturbation theory~\cite{Weinberg:1990rz,Weinberg:1991um}. It has been remarkably successful and entered into a era of high precision since the year of 2003~\cite{Entem:2003ft,Epelbaum:2004fk,Epelbaum:2008ga,Machleidt:2011zz,Epelbaum:2014sza,Entem:2015xwa}. At the core of the chiral nuclear force is the contact nucleon-nucleon chiral Lagrangian, which parameterizes
the short-range interactions not captured by the long-range pion exchanges. In the present work, we would like to study the constraints of relativity/Lorentz covariance on
the chiral NN contact Lagrangian. Up to now, there are already a few works in this direction. In Refs.~\cite{Girlanda:2010ya,Girlanda2011}, the relativistic $NN$ Lagrangian is constructed up to  $\mathcal{O}(q^2)$. In
Ref.~\cite{Petschauer:2013uua}, such a work is extended to the SU(3) case. In a series of recent works,  using the leading order relativistic chiral Lagrangian,
some of us have studied nucleon-nucleon scattering data~\cite{Ren:2016jna} and hyperon-nucleon (hyperon) data~\cite{Li:2016mln} as well as related lattice QCD simulations~\cite{Song:2018qqm,Li:2018tbt}.
Quite lately, it was shown in Ref.~\cite{Girlanda:2018xrw} that  one can achieve 
 a satisfactory description of the polarized $p$-$d$ scattering data below the deuteron breakup threshold either with the leading order relativistic $3N$  interaction (in
 the power counting of Ref.~\cite{Ren:2016jna}) or with the next-to-leading order non-relativistic $3N$ interaction. 
 In particular, the long-standing $A_y$ puzzle seems to be solved .

In this work, we revisit this problem in the $NN$ sector and construct a complete set of covariant $NN$ contact Lagrangian terms up to $\mathcal{O}(q^4)$.  Using
 the equation of motion to eliminate redundant terms, the total number of independent terms are found to be 4 at  $\mathcal{O}(q^0)$,  13 at $\mathcal{O}(q^2)$, and 23 at $\mathcal{O}(q^4)$. As
an explicit check, we expand our relativistic~\footnote{In the present work,  ``relativistic'' and ``covariant" are used interchangeably.} Lagrangian in terms of $1/m_N$, where $m_N$ is the nucleon mass, and show that in the heavy baryon limit our Lagrangian reduces to that of the HB$\chi$PT.

This work is
organized as follows. In Sec. II, we explain the general principles to construct a covariant chiral nucleon-nucleon Lagrangian, list the linear relations derived from the equation of motion and use them to eliminate redundant terms,
 and write down the covariant nucleon-nucleon contact Lagrangian up to order $\mathcal{O}(q^4)$.
In Sec. III, we compare our results with those of earlier studies and perform the non-relativistic reduction, followed by a short summary in Sec. IV. 
A concise derivation of the non-relativistic nucleon-nucleon contact Lagrangian up to order $\mathcal{O}(q^4)$ is given in the Appendix.

\section{Relativistic chiral $NN$ Lagrangians up to $\mathcal{O}(q^4)$ }
The relativistic chiral $NN$ contact Lagrangian should fulfill the following requirements. First of all, the Lagrangian must be a Lorentz scalar. Secondly, it has to be invariant under chiral, parity ($\mathcal{P}$), charge conjugation ($\mathcal{C}$), hermitian conjugation (h.c.), and time reversal transformations. Thirdly,  it needs to satisfy a proper power counting  so that one can determine the relative importance of each term and
have limited terms at each order in the chiral expansion.

The general expression of a covariant nucleon-nucleon contact Lagrangian~\footnote{In the present work, to simplify the derivation,
we do not consider the interaction of external fields and the pion with the nucleon.} reads,
\begin{eqnarray}\label{eq:expression lag}
\frac{1}{\left(2m\right)^{N_d}} \left(\bar{\psi} i \overleftrightarrow {\partial}^{\alpha} i\overleftrightarrow {\partial}^{\beta} ... \Gamma_A \psi\right)
 \partial^{\lambda} \partial^{\mu} ... \left(\bar{\psi} i \overleftrightarrow {\partial}^{\sigma} i\overleftrightarrow {\partial}^{\tau} ... \Gamma_B \psi\right),
\end{eqnarray}
where $\psi$~\footnote{$\psi=(\psi_p,\psi_n)^T$. In the present work,  since we work in the isospin limit, we do not explicitly specify the fact that  $\psi$ is an isospin doublet. The same applies to the
non-relativistic nucleon field $N$. } and $\bar{\psi}$ donate the relativistic nucleon  field, $\overleftrightarrow {\partial}^{\alpha} = \overrightarrow{\partial}^{\alpha} - \overleftarrow{\partial}^{\alpha}$, 
where $\overrightarrow{\partial}^\alpha /\overleftarrow{\partial}^\alpha$ refers to the derivative on $\psi/ \bar{\psi}$, and $\Gamma \in \{\mathbbm{1}, \gamma_5, \gamma^{\mu}, \gamma_5 \gamma^{\mu}, \sigma^{\mu\nu},g^{\mu\nu}, \epsilon^{\mu\nu\rho\sigma}\} $.  In the above equation, $N_d$ refers to the number of four-derivatives (both $\overleftrightarrow \partial $ and $\partial$ ) in the Lagrangian, $m$ refers to the nucleon mass in the chiral limit, and the factor $1/(2m)^{N_{d}}$ has been introduced, so that all the contact terms have the same dimension~\cite{Girlanda:2010ya}.

For the construction of effective Lagrangians, symmetry constraints are the most important. In our present case, apart from the invariance under Lorentz transformation, the covariant nucleon-nucleon contact Lagrangian has to be invariant under local chiral, parity, charge conjugation, hermitian conjugation, and time reversal transformations. The Lorentz indices $\alpha, \beta...$ have to be contracted among themselves to fulfill  Lorentz invariance. The local chiral symmetry can be trivially  fulfilled because the nucleon field $\psi$ transforms under chiral rotation $\psi\rightarrow K \psi$ and $K K^\dagger = \mathbbm{1}$, where $K\in \mathrm{SU(2)_V}$. The hermitian conjugation symmetry does not impose any constraint since to fulfill it one can always multiply the Lagrangian with a proper choice of factor $i$. According to the CPT  theorem, time reversal symmetry is also automatically fulfilled if  parity symmetry and charge conjugation symmetry are fulfilled. Therefore, one only needs to make sure that parity and charge conjugation symmetries  are fulfilled.

\begin{table}[h]
\caption{Chiral dimensions and properties of fermion bilinears, derivative operators,  Dirac matrices, and Levi-Civita tensor, under parity ($\mathcal{P}$), charge conjugation ($\mathcal{C}$), and hermitian conjugation (h.c.)  transformations.}
\label{Nucleon bilinears}
\centering
\begin{tabular}{ccccccccc}
 \hline
 \hline
 &  $\mathbbm{1}$    &  $\gamma_5$    &    $\gamma_\mu$     &   $\gamma_5\gamma_\mu$   &   $\sigma_{\mu\nu}$  &   $\epsilon_{\mu\nu\rho\sigma}$ &   $\overleftrightarrow \partial_{\mu}$ & $\partial_{\mu}$\\
 \hline
   $\mathcal{O}$   & $0$ & $1$ & $0$ & $0$ & $0$ & $-$ & $0$ & $1$\\
  $\mathcal{P}$   & $+$ & $-$ & $+$ & $-$ & $+$ & $-$ & $+$ & $+$\\
  $\mathcal{C}$   & $+$ & $+$ & $-$ & $+$ & $-$ & $+$ & $-$ & $+$\\
  h.c.            & $+$ & $-$ & $+$ & $+$ & $+$ & $+$ & $-$ & $+$\\

  \hline
  \hline
\end{tabular}
\end{table}

To construct the chiral Lagrangian, one has to specify a proper power counting. In our present case, we need to specify the chiral dimensions of
all the building blocks. In the covariant case, the power counting is more involved, compared to the non-relativistic case (see the Appendix).
The chiral dimensions and properties of fermion bilinears, derivative operators, Dirac matrices, and Levi-Civita tensor under parity, charge conjugation, and hermitian conjugation transformations are listed in Table~\ref{Nucleon bilinears}. The derivative $\partial$ acting on the whole bilinear is of order $\mathcal{O}(q^1)$, while the derivative $\overleftrightarrow {\partial} $ acting inside a bilinear is of $\mathcal{O}(q^0)$ due to the presence of the nucleon mass,
where $q$ denotes a genetic small quantity, such as the nucleon three momentum or the pion mass. The Dirac matrix $\gamma_5$ is of order $\mathcal{O}(q^1)$ because it mixes the large and small components of the Dirac spinor. The Levi-Civita tensor $\epsilon_{\mu \nu \rho \sigma}$ contracting with $n$ derivatives acting inside a bilinear raises the chiral order by $n-1$. If a derivative $\overleftrightarrow \partial$ is contracted with one of the Dirac matrices $\gamma_5\gamma^{\mu}$ or $\sigma^{\mu\nu}$  in a different bilinear, the matrix element is of $\mathcal{O}(q^1)$,
as can be explicitly checked by means of the equation of motion (EOM). Therefore, at each order in the powering counting, only a finite number of $\partial$ and $\epsilon_{\mu\nu\rho\sigma}$ appear. However, in principle, any numbers of pairwise contracted $i\overleftrightarrow{\partial}$ of the form
\begin{equation}\label{eq:npartial}
\widetilde{\mathcal{O}}_{\Gamma_A \Gamma_B}^{(n)}=\frac{1}{(2m)^{2n}}\left(\bar{\psi} i \overleftrightarrow{\partial}^{\mu_1} i \overleftrightarrow{\partial}^{\mu_2} ... i \overleftrightarrow{\partial}^{\mu_n} \Gamma_{A}^{\alpha}\psi\right)
 \times\left(\bar{\psi} i \overleftrightarrow{\partial}_{\mu_1} i \overleftrightarrow{\partial}_{\mu_2} ... i \overleftrightarrow{\partial}_{\mu_n} \Gamma_{B \alpha}\psi\right),
\end{equation}
 is allowed, since it is of  $\mathcal{O}(q^0)$.  On the other hand,
such a structure
\begin{equation}
\frac{\left[\left(p_1+p_3\right) \cdot \left(p_2+p_4\right)\right]^{n}}{\left(2m\right)^{2n}},
\end{equation}
can be rewritten as
\begin{eqnarray}
\left[1+\frac{\left(s-4m^2\right)-u}{4m^2}\right]^{n},
\end{eqnarray}
with $s-4m^2=-(p_1-p_2)^2=-(p_3-p_4)^2 \sim \mathcal{O}(q^2)$ and $u=(p_1-p_4)^2\sim \mathcal{O}(q^2)$. ~\footnote{This can be easily checked by noting that the two incoming (outgoing) nucleons should be equally off-shell
in the CM frame.}
Therefore, at $\mathcal{O}(q^0)$, only the terms with $n=0,1,2$ are needed, at  $\mathcal{O}(q^2)$ only the terms with $n=0,1$ are needed, and at $\mathcal{O}(q^4)$ only the terms with $n=0$ are needed since
no new structures appear for $n$ larger than those specified above.

Following the general principles of constructing effective Lagrangians and guided by Table~\ref{Nucleon bilinears}, one can 
write down all the terms of  $\mathcal{O}(q^0)$, $\mathcal{O}(q^2)$, and $\mathcal{O}(q^4)$. As we show in the following, not all of them are independent up to the order of our concern
and one can use the EOM  to eliminate the non-independent or redundant terms.

%\subsection{ Reducing redundant terms  using  equation of motion}

\begin{table*}[!h]
    \caption{Decomposition of the Dirac matrix products \(\Gamma\gamma_\lambda\) into charge conjugation even and charge conjugation odd parts~\cite{Petschauer:2013uua}.}
  \label{tab:gammadec}
  \centering
  \begin{tabular}{ccc}
  \hline
  \hline
  $\Gamma$             &    $\Gamma_\lambda^{'}$     & $\Gamma_\lambda^{''}$\\
  \hline
  $\mathbbm{1}$        &    $\gamma_\lambda$         &   $0$\\
  $\gamma_\mu$         &    $g_{\mu\lambda} 1$       &   $-i\sigma_{\mu\lambda}$\\
  $\gamma_5$           &    $0$                      &   $\gamma_5 \gamma_\lambda$\\
  $\gamma_5\gamma_\mu$ &    $\frac{1}{2}\epsilon_{\mu\lambda \rho\tau}\sigma^{\rho\tau}$   &$g_{\mu\lambda}\gamma_5$\\
  $\sigma_{\mu\nu}$    &    $\epsilon_{\mu\nu\lambda\tau}\gamma_5\gamma^\tau$
  &  $-i\left(g_{\mu\lambda}\gamma_\nu-g_{\nu\lambda}\gamma_\mu\right)$\\
  \hline
    $\epsilon_{\mu\nu\rho\tau}\gamma^\tau$
  & $\epsilon_{\mu\nu\rho\lambda}1$
  & $g_{\mu\lambda}\gamma_5\sigma_{\nu\rho}+g_{\rho\lambda}\gamma_5 \sigma_{\mu\nu}
  +g_{\nu\lambda}\gamma_5\sigma_{\rho \mu}$\\
    $\epsilon_{\mu\nu\rho\tau}\gamma_5\gamma^\tau$
  & $g_{\mu\lambda}\sigma_{\nu\rho}+ g_{\rho\lambda}\sigma_{\mu\nu}+g_{\nu\lambda}\sigma_{\rho \mu}$
  & $\epsilon_{\mu\nu\rho\lambda}\gamma_5$\\
    $\epsilon_{\mu\nu\rho\alpha} \sigma^{\alpha}_\tau$
  & $\gamma_5\gamma_\rho\left(g_{\lambda\nu}g_{\mu\tau}- g_{\lambda\mu}g_{\nu\tau}\right)
     +\gamma_5\gamma_\nu \left(g_{\lambda\mu}g_{\rho\tau}-g_{\lambda\rho}g_{\mu\tau} \right)
     +\gamma_5\gamma_\mu \left(g_{\lambda\rho}g_{\nu\tau}-g_{\lambda\nu}g_{\rho\tau}\right)$
  & $ig_{\lambda\tau}\epsilon_{\mu\nu\rho\alpha} \gamma^\alpha-i\epsilon_{\mu\nu\rho\lambda}\gamma_{\tau}$\\
    $\frac{i}{2}\epsilon_{\mu\nu\rho\tau} \sigma^{\rho\tau}=\gamma_5\sigma_{\mu\nu}$
  & $\frac{1}{i}\left(g_{\mu\lambda}\gamma_5 \gamma_\nu-g_{\nu\lambda}\gamma_5\gamma_\mu\right)$
  & $\epsilon_{\mu\nu\lambda\rho}\gamma^\rho$\\
  \hline
  \hline
  \end{tabular}
  \end{table*}
The equation of motion for the nucleon refers to the well-known Dirac equation at LO
\begin{equation}
 \slashed{\partial}\psi = \gamma^\mu \partial_\mu \psi = - i m \psi + \mathcal{O}(q) \,,
\end{equation}
and its Hermitian conjugate. Up to higher order corrections one can replace \(\slashed{\partial}\psi\) by \(- i m \psi\) and \(\bar \psi \overleftarrow{\slashed{\partial}}\) by \( i m \bar \psi\).
To fully utilize this EOM,  one needs to transform terms that do not contain $\slashed{\partial}$ into forms that contain it. Such a technique has been extensively discussed
in the construction of the $\pi N$ Lagrangian~\cite{Fettes:2000gb} and baryon-baryon Lagrangian~\cite{Petschauer:2013uua}.
The details can be found in Refs.~\cite{Fettes:2000gb,Petschauer:2013uua}. The master  formula is

{
\begin{equation} \begin{aligned} \label{eq:eomrel}
 - 2i m \left(\bar{\psi} {\Gamma} \psi\right)&\approx  \left(\bar{\psi} {\Gamma^\prime}^\lambda \overleftrightarrow{\partial}_{\lambda}\psi\right)+\partial_{\lambda}\left(\bar{\psi} {\Gamma^{\prime\prime}}^\lambda  \psi\right) \,,
\end{aligned} \end{equation}
}
where $\Gamma$, $\Gamma^{\prime}$, and $\Gamma^{\prime\prime}$ are Dirac matrices listed in Table~\ref{tab:gammadec} and $\approx$ indicates equal up to higher orders.
Using the EOM together with the decomposition of Dirac matrices, one can obtain the following linear relations:
\begin{subequations}\label{eq:linear relations}
\renewcommand{\theequation}
{\theparentequation.\arabic{equation}}{
\begin{align}
& \partial^\mu\left( \bar{\psi} \gamma_{\mu} \psi \right) \approx 0,\\
& \partial^\mu\left( \bar{\psi} \overleftrightarrow{\partial}_\mu \psi\right) \approx 0,\\
&\partial^{\mu} \left(\bar {\psi} \gamma_5 \gamma_\mu \psi \right) \approx -2m \left(\bar {\psi}i \gamma_5  \psi\right),\\
&\partial^{\mu} \left(\bar {\psi} \sigma_{\mu\nu} \psi \right)  \approx \left(\bar{\psi}  i\overleftrightarrow {\partial}_\nu \psi\right)- 2m \left(\bar{\psi} \gamma_\nu  \psi\right) ,\\
&\left( \bar{\psi} \gamma^{\mu} i\overleftrightarrow\partial_\mu \psi \right) \approx 2m \left( \bar{\psi} \psi \right),\\
&\left(\bar{\psi} \overleftrightarrow{\partial}^2\psi\right) \approx -4m^2\left(\bar{\psi} \psi\right) - \partial^{2}\left(\bar{\psi} \psi\right),\\
&\left(\bar {\psi} \gamma_5 \gamma^\mu i\overleftrightarrow\partial_\mu\psi \right) \approx 0,\\
&\left(\bar{\psi}i \overleftrightarrow \partial_{\mu} \sigma^{\mu\nu} \psi \right) \approx -\partial^\nu \left(\bar{\psi} \psi\right),\\
&-2i m \left(\bar{\psi} \gamma_5 \gamma^\mu \psi\right) \approx \left( \bar \psi \frac{1}{2} \epsilon^{\mu\lambda\rho\tau}\sigma_{\rho\tau}  \overleftrightarrow{\partial}_\lambda \psi\right) +  \partial^{\mu}\left(\bar{\psi} \gamma_5 \psi \right),\\
&-2i m \left(\bar{\psi} \sigma^{\mu\nu} \psi\right) \approx \left( \bar \psi \epsilon^{\mu\lambda\rho\tau}\gamma_5 \gamma_{\tau} \overleftrightarrow{\partial}_\lambda \psi\right)+ \partial_{\lambda}\left(\bar{\psi} -i \left(g^{\mu\lambda}\gamma^{\nu}-g^{\nu\lambda} \gamma^{\mu}  \right) \psi \right),\\
&-2im\left(\bar{\psi} \epsilon_{\mu\nu\rho\tau} \gamma^{\tau}\psi\right) \approx \left(\bar{\psi} \epsilon_{\mu\nu\rho\lambda} \overleftrightarrow{\partial}^{\lambda} \psi\right) + \partial^{\lambda}\left(\bar{\psi} \left( g_{\mu\lambda}\gamma_5\sigma_{\nu\rho}+g_{\rho\lambda}\gamma_5\sigma_{\mu\nu} +g_{\nu\lambda}\gamma_5\sigma_{\rho\mu} \right) \psi \right),\\
&-2im \left( \bar{\psi} \epsilon_{\mu\nu\rho\tau} \gamma_5\gamma^{\tau} \psi \right) \approx \left( \bar{\psi} \left( \sigma_{\nu\rho} \overleftrightarrow{\partial}_{\mu}+ \sigma_{\mu\nu} \overleftrightarrow{\partial}_{\rho}+  \sigma_{\rho\mu} \overleftrightarrow{\partial}_{\nu}\right) \psi \right) +\partial^{\lambda}\left(\bar{\psi}\epsilon_{\mu\nu\rho\lambda} \gamma_5 \psi\right),\\
&-2im\left( \bar{\psi} \epsilon_{\mu\nu\rho\alpha} {\sigma^{\alpha}}_{\tau} \psi \right)  \approx \left( \bar{\psi} \left( \gamma_5\gamma_\rho \left(\overleftrightarrow{\partial}_{\nu}g_{\mu\tau}- \overleftrightarrow{\partial}_{\mu}g_{\nu\tau}\right)+\gamma_5\gamma_{\nu}\left( \overleftrightarrow{\partial}_{\mu} g_{\rho\tau}-\overleftrightarrow{\partial}_{\rho}g_{\mu\tau}\right)+\gamma_5\gamma_{\mu} \left(\overleftrightarrow{\partial}_{\rho}g_{\nu\tau}-\overleftrightarrow{\partial}_{\nu}g_{\rho\tau} \right)\right)\psi\right)\\\nonumber
&+\partial^{\lambda}\left( \bar{\psi}\left( ig_{\lambda\tau}\epsilon_{\mu\nu\rho\alpha} \gamma^{\alpha}-i\epsilon_{\mu\nu\rho\lambda}\gamma_{\tau} \right) \psi\right),\\
&m\left( \bar{\psi} \epsilon_{\mu\nu\rho\tau} \sigma^{\rho \tau} \psi \right)  \approx \left( \bar{\psi} \left(\gamma_5\gamma_\mu i\overleftrightarrow{\partial}_{\nu}- \gamma_5\gamma_\nu i\overleftrightarrow{\partial}_{\mu}\right)\psi\right)+\partial^{\lambda}\left(\bar{\psi}\epsilon_{\mu\nu\lambda\rho} \gamma^{\rho} \psi\right),\\
&\left(\bar{\psi} \epsilon_{\mu\nu\alpha\beta} i\overleftrightarrow{\partial}^{\mu} i\overleftrightarrow{\partial}^{\nu}  \cdots \psi\right)=0,\\
&\epsilon_{\mu\nu\alpha\beta}   \partial^{\mu} \partial^{\nu} \left(\bar{\psi} \cdots \psi\right)=0.
\end{align}
}
\end{subequations}
The set of relations in Eq.~\eqref{eq:linear relations} lead to the following four simplification  rules:
\begin{enumerate}
\item Terms with $\epsilon_{\mu\nu\rho\tau}$ can always be transformed into those without it, so no terms with $\epsilon_{\mu\nu\rho\tau}$ are needed,
\item The derivative $\partial_{\mu}$ acting on the whole fermion bilinear cannot be contracted with any elements of the Clifford algebra except for $\sigma^{\mu\nu}$,
\item The derivative $i \overleftrightarrow{\partial}^{\mu}$ cannot be contracted with any elements of the Clifford algebra inside the same fermion bilinear,
\item Terms with $\gamma^{\mu}$ can be transformed into terms with $i \overleftrightarrow{\partial}^{\mu}$ except for the cases where it is contracted with $\sigma^{\mu\nu}$.
\end{enumerate}

\begin{table}[!h]
\caption{A complete set of relativistic $NN$ contact  Lagrangian up to $\mathcal{O}(q^4)$. } \label{tb:NNLO NN Lagrangian}
\centering
\begin{tabular}{c|c|c|c}
   \hline
   \hline
  $\widetilde{{O}}_{1}$          &$\left(\bar{\psi} \psi\right)\left(\bar{\psi} \psi\right)$   &$\widetilde{{O}}_{21}$       &$\frac{1}{16m^4}\left(\bar{\psi} i\overleftrightarrow{\partial}^{\mu} \psi\right) \partial^{2} \partial^{\nu}\left(\bar{\psi} \sigma_{\mu\nu} \psi\right)$    \\
  $\widetilde{{O}}_{2}$          &$\left(\bar{\psi} \gamma^{\mu} \psi\right)\left(\bar{\psi} \gamma_{\mu} \psi\right)$  & $\widetilde{{O}}_{22}$       &$\frac{1}{16m^4}\left(\bar{\psi} \sigma^{\mu \alpha} \psi\right) \partial^{2} \partial_{\alpha}\partial^{\nu}\left(\bar{\psi} \sigma_{\mu\nu} \psi\right)$  \\
  $\widetilde{{O}}_{3}$          &  $\left(\bar{\psi} \gamma_5 \gamma^{\mu} \psi\right)\left(\bar{\psi} \gamma_5 \gamma_{\mu} \psi\right)$    & $\widetilde{{O}}_{23}$&  $\frac{1}{16m^4}\left(\bar{\psi} \sigma^{\mu\nu} i\overleftrightarrow{\partial}^{\alpha} \psi\right) \partial^{\beta} \partial_{\nu}\left(\bar{\psi} \sigma_{\alpha \beta} i\overleftrightarrow{\partial}_{\mu} \psi\right)$ \\
  $\widetilde{{O}}_{4}$          & $\left(\bar{\psi} \sigma^{\mu\nu} \psi\right)\left(\bar{\psi} \sigma_{\mu\nu} \psi\right)$ & $\widetilde{{O}}_{24}$       &$\frac{1}{16m^4}\left(\bar{\psi} \psi\right) \partial^{4}\left(\bar{\psi} \psi\right)$\\
  \cline{1-2}
  $\widetilde{{O}}_{5}$          & $\left(\bar{\psi} \gamma_5 \psi\right)\left(\bar{\psi} \gamma_5 \psi\right)$ & $\widetilde{{O}}_{25}$       &$\frac{1}{16m^4}\left(\bar{\psi} \gamma^{\mu} \psi\right) \partial^{4} \left(\bar{\psi} \gamma_{\mu} \psi\right)$    \\
  $\widetilde{{O}}_{6}$          &$\frac{1}{4m^2}\left(\bar{\psi} \gamma_5 \gamma^{\mu} i\overleftrightarrow{\partial}^{\alpha} \psi\right)\left(\bar{\psi} \gamma_5 \gamma_{\alpha} i \overleftrightarrow{\partial}_{\mu} \psi\right)$   & $\widetilde{\mathcal{O}}_{26}$       &$\frac{1}{16m^4}\left(\bar{\psi} \gamma_5 \gamma^{\mu} \psi\right) \partial^{4} \left(\bar{\psi} \gamma_5 \gamma_{\mu} \psi\right)$  \\
  $\widetilde{{O}}_{7}$          &$\frac{1}{4m^2}\left(\bar{\psi} \sigma^{\mu\nu} i\overleftrightarrow{\partial}^{\alpha} \psi\right)\left(\bar{\psi} \sigma_{\mu\alpha} i\overleftrightarrow{\partial}_{\nu} \psi\right)$ & $\widetilde{{O}}_{27}$       &$\frac{1}{16m^4}\left(\bar{\psi} \sigma^{\mu\nu} \psi\right) \partial^{4} \left(\bar{\psi} \sigma_{\mu\nu} \psi\right)$ \\
  $\widetilde{{O}}_{8}$          &$\frac{1}{4m^2}\left(\bar{\psi}  i\overleftrightarrow{\partial}^{\mu} \psi\right)\partial^{\nu}\left(\bar{\psi} \sigma_{\mu\nu} \psi\right)$ & $\widetilde{{O}}_{28}$  &   $\frac{1}{4m^2}\left(\bar{\psi} \gamma_5 i\overleftrightarrow{\partial}^{\alpha} \psi\right)\left(\bar{\psi} \gamma_5 i\overleftrightarrow{\partial}_{\alpha} \psi\right)-\widetilde{O}_{5}$\\
  $\widetilde{{O}}_{9}$          &$\frac{1}{4m^2}\left(\bar{\psi}  \sigma^{\mu \alpha} \psi\right)\partial_{\alpha}\partial^{\nu}\left(\bar{\psi} \sigma_{\mu\nu} \psi\right)$   & $\widetilde{{O}}_{29}$  & $\frac{1}{16m^4}\left(\bar{\psi} \gamma_5\gamma^{\mu} i\overleftrightarrow{\partial}^{\alpha} i\overleftrightarrow{\partial}^{\beta} \psi\right)\left(\bar{\psi} \gamma_5\gamma_{\alpha} i\overleftrightarrow{\partial}_{\mu} i\overleftrightarrow{\partial}_{\beta} \psi\right)-\widetilde{O}_{6}$\\
  $\widetilde{{O}}_{10}$  &$\frac{1}{4m^2}\left(\bar{\psi} \psi\right) \partial^{2}\left(\bar{\psi} \psi\right)$  & $\widetilde{{O}}_{30}$  & $\frac{1}{16m^4}\left(\bar{\psi} \sigma^{\mu\nu} i\overleftrightarrow{\partial}^{\alpha} i\overleftrightarrow{\partial}^{\beta} \psi\right)\left(\bar{\psi} \sigma_{\mu\alpha} i\overleftrightarrow{\partial}_{\nu} i\overleftrightarrow{\partial}_{\beta} \psi\right)-\widetilde{O}_{7}$ \\
  $\widetilde{{O}}_{11}$  &$\frac{1}{4m^2}\left(\bar{\psi} \gamma^{\mu} \psi\right) \partial^{2} \left(\bar{\psi} \gamma_{\mu} \psi\right)$  & $\widetilde{{O}}_{31}$  & $\frac{1}{16m^4}\left(\bar{\psi}  i\overleftrightarrow{\partial}^{\mu} i\overleftrightarrow{\partial}^{\beta} \psi\right) \partial^{\alpha}\left(\bar{\psi} \sigma_{\mu\alpha}  i\overleftrightarrow{\partial}_{\beta} \psi\right)-\widetilde{O}_{8}$\\
  $\widetilde{{O}}_{12}$       &$\frac{1}{4m^2}\left(\bar{\psi} \gamma_5 \gamma^{\mu} \psi\right) \partial^{2} \left(\bar{\psi} \gamma_5 \gamma_{\mu} \psi\right)$   & $\widetilde{{O}}_{32}$  & $\frac{1}{16m^4}\left(\bar{\psi}  \sigma^{\mu \alpha} i\overleftrightarrow{\partial}^{\beta} \psi\right)\partial_{\alpha}  \partial^{\nu}\left(\bar{\psi} \sigma_{\mu\nu}  i\overleftrightarrow{\partial}_{\beta} \psi\right)-\widetilde{O}_{9}$ \\
  $\widetilde{{O}}_{13}$       &$\frac{1}{4m^2}\left(\bar{\psi} \sigma^{\mu\nu} \psi\right) \partial^{2} \left(\bar{\psi} \sigma_{\mu\nu} \psi\right)$  &  $\widetilde{{O}}_{33}$ &  $\frac{1}{16m^4}\left(\bar{\psi} i\overleftrightarrow{\partial}^{\alpha} \psi\right)\partial^{2}\left(\bar{\psi} i\overleftrightarrow{\partial}_{\alpha} \psi\right)-\widetilde{O}_{10}$ \\
  $\widetilde{{O}}_{14}$  & $\frac{1}{4m^2}\left(\bar{\psi} i\overleftrightarrow{\partial}^{\alpha} \psi\right)\left(\bar{\psi} i\overleftrightarrow{\partial}_{\alpha} \psi\right)-\widetilde{O}_{1}$    & $\widetilde{{O}}_{34}$ &  $\frac{1}{16m^4}\left(\bar{\psi} \gamma^{\mu} i\overleftrightarrow{\partial}^{\alpha} \psi\right)\partial^{2}\left(\bar{\psi} \gamma_{\mu} i\overleftrightarrow{\partial}_{\alpha} \psi\right)-\widetilde{O}_{11}$ \\
  $\widetilde{{O}}_{15}$  & $\frac{1}{4m^2}\left(\bar{\psi} \gamma^{\mu} i\overleftrightarrow{\partial}^{\alpha} \psi\right)\left(\bar{\psi} \gamma_{\mu} i\overleftrightarrow{\partial}_{\alpha} \psi\right)-\widetilde{O}_{2}$   &$\widetilde{{O}}_{35}$ &  $\frac{1}{16m^4}\left(\bar{\psi} \gamma_5\gamma^{\mu} i\overleftrightarrow{\partial}^{\alpha} \psi\right)\partial^{2}\left(\bar{\psi} \gamma_5\gamma_{\mu} i\overleftrightarrow{\partial}_{\alpha} \psi\right)-\widetilde{O}_{12}$  \\
   $\widetilde{{O}}_{16}$  &   $\frac{1}{4m^2}\left(\bar{\psi} \gamma_5\gamma^{\mu} i\overleftrightarrow{\partial}^{\alpha} \psi\right)\left(\bar{\psi} \gamma_5 \gamma_{\mu} i\overleftrightarrow{\partial}_{\alpha} \psi\right)-\widetilde{O}_{3}$    &  $\widetilde{{O}}_{36}$ &  $\frac{1}{16m^4}\left(\bar{\psi} \sigma^{\mu\nu} i\overleftrightarrow{\partial}^{\alpha} \psi\right)\partial^{2}\left(\bar{\psi} \sigma_{\mu\nu} i\overleftrightarrow{\partial}_{\alpha} \psi\right)-\widetilde{O}_{13}$ \\
  $\widetilde{{O}}_{17}$  &   $\frac{1}{4m^2}\left(\bar{\psi} \sigma^{\mu\nu} i\overleftrightarrow{\partial}^{\alpha} \psi\right)\left(\bar{\psi} \sigma_{\mu\nu} i\overleftrightarrow{\partial}_{\alpha} \psi\right)-\widetilde{O}_{4}$  & $\widetilde{{O}}_{37}$ & $\frac{1}{16m^4}\left(\bar{\psi} i\overleftrightarrow{\partial}^{\alpha} i\overleftrightarrow{\partial}^{\beta} \psi\right)\left(\bar{\psi} i\overleftrightarrow{\partial}_{\alpha} i\overleftrightarrow{\partial}_{\beta} \psi\right)-2\widetilde{O}_{14}-\widetilde{O}_{1}$ \\
  \cline{1-2}
  $\widetilde{{O}}_{18}$       &$\frac{1}{4m^2}\left(\bar{\psi} \gamma_5 \psi\right) \partial^{2}\left(\bar{\psi} \gamma_5 \psi\right)$  &$\widetilde{{O}}_{38}$ & $\frac{1}{16m^4}\left(\bar{\psi} \gamma^{\mu} i\overleftrightarrow{\partial}^{\alpha} i\overleftrightarrow{\partial}^{\beta} \psi\right)\left(\bar{\psi} \gamma_{\mu} i\overleftrightarrow{\partial}_{\alpha} i\overleftrightarrow{\partial}_{\beta} \psi\right)-2\widetilde{O}_{15}-\widetilde{O}_{2}$ \\
  $\widetilde{{O}}_{19}$       &$\frac{1}{16m^4}\left(\bar{\psi} \gamma_5 \gamma^{\mu} i\overleftrightarrow{\partial}^{\nu} \psi\right) \partial^{2} \left(\bar{\psi} \gamma_5 \gamma_{\nu} i\overleftrightarrow{\partial}_{\mu}  \psi\right)$ &   $\widetilde{{O}}_{39}$ & $\frac{1}{16m^4}\left(\bar{\psi} \gamma_5 \gamma^{\mu} i\overleftrightarrow{\partial}^{\alpha} i\overleftrightarrow{\partial}^{\beta} \psi\right)\left(\bar{\psi} \gamma_5 \gamma_{\mu} i\overleftrightarrow{\partial}_{\alpha} i\overleftrightarrow{\partial}_{\beta} \psi\right)-2\widetilde{O}_{16}-\widetilde{O}_{3}$ \\
  $\widetilde{{O}}_{20}$       &$\frac{1}{16m^4}\left(\bar{\psi} \sigma^{\mu\nu} i\overleftrightarrow{\partial}^{\alpha} \psi\right) \partial^{2} \left(\bar{\psi} \sigma_{\mu\alpha} i\overleftrightarrow{\partial}_{\nu}  \psi\right)$ &  $\widetilde{{O}}_{40}$ & $\frac{1}{16m^4}\left(\bar{\psi} \sigma^{\mu\nu} i\overleftrightarrow{\partial}^{\alpha} i\overleftrightarrow{\partial}^{\beta} \psi\right)\left(\bar{\psi} \sigma_{\mu\nu} i\overleftrightarrow{\partial}_{\alpha} i\overleftrightarrow{\partial}_{\beta} \psi\right)-2\widetilde{O}_{17}-\widetilde{O}_{4}$  \\
   \hline
   \hline
    \end{tabular}
\end{table}

Using the four rules listed above, we obtain a minimal and complete set of relativistic $NN$ contact Lagrangian terms, which are summarized in Table~\ref{tb:NNLO NN Lagrangian}. The Lagrangian up to order $\mathcal{O}(q^4)$ now contains  %39
40 terms which fulfill the power counting specified above.

\section{Discussions}

\subsection{Comparison with previous works}

 The Lagrangian listed in Table~\ref{tb:NNLO NN Lagrangian} is different from those of Refs.~\cite{Girlanda:2010ya,Girlanda2011,Petschauer:2013uua}. Moreover, the results of Refs.~\cite{Girlanda:2010ya,Girlanda2011,Petschauer:2013uua} can be reduced to ours using the linear relations listed in the last section. Compared with the results of Refs.~\cite{Girlanda:2010ya,Girlanda2011},  in our case all the terms with $\epsilon_{\mu\nu\alpha\beta}$ are eliminated  using the EOM. In Ref.~\cite{Petschauer:2013uua}, their results contain at most two pairwise contracted $i\overleftrightarrow{\partial}$ of the form $\frac{1}{(2m)^{2n}}\left(\bar{\psi} i \overleftrightarrow{\partial}^{\mu_1} i \overleftrightarrow{\partial}^{\mu_2} ... i \overleftrightarrow{\partial}^{\mu_n} \Gamma_{A}^{\alpha}\psi\right)\left(\bar{\psi} i \overleftrightarrow{\partial}_{\mu_1} i \overleftrightarrow{\partial}_{\mu_2} ... i \overleftrightarrow{\partial}_{\mu_n} \Gamma_{B \alpha}\psi\right)$ up to  $\mathcal{O}(q^2)$ while we only include at most one, which is consistent with  Ref.~\cite{Girlanda:2010ya}. We note that the terms with $\partial^{\mu}\left(\bar{\psi} \sigma_{\mu\nu} \psi\right)$ are included in our Lagrangian but are not in those of Refs.~\cite{Girlanda:2010ya,Girlanda2011,Petschauer:2013uua}. This is just a different choice of independent terms,  because $\partial^{\mu}\left(\bar{\psi} \sigma_{\mu\nu} \psi\right)=\left(\bar{\psi} \sigma_{\mu\nu} \overrightarrow{\partial}^{\mu}\psi\right)+\left(\bar{\psi}\overleftarrow{\partial}^{\mu} \sigma_{\mu\nu} \psi\right) \approx \left(\bar{\psi} i \overleftrightarrow{\partial}_{\nu} \psi\right)-2m\left(\bar{\psi} \gamma_{\nu} \psi\right)$. In Refs.~\cite{Girlanda:2010ya,Girlanda2011},  { 
 {terms containing $\left(\bar{\psi} i\overleftrightarrow{\partial}^{\mu} \psi\right)\left(\bar{\psi} \gamma_{\mu} \psi\right)$ are included, while in our case, we replace them with $\left( \bar{\psi} i\overleftrightarrow{\partial}^{\mu} \psi\right) \partial^{\nu}\left( \bar{\psi} \sigma_{\mu\nu} \psi\right)$} }using the EOM. Note further that in Ref.~\cite{Petschauer:2013uua}, the terms with $\partial^{\mu}\left(\bar{\psi} \sigma_{\mu\nu} \psi\right)$ are not included because such terms are argued to be of higher order.
 However, we prefer to keep these terms because they are of unique Lorentz structure which satisfy our power counting rules. The $\Gamma_{i}$'s in our work are also different from those in Ref.~\cite{Petschauer:2013uua}. In our work, we take $\Gamma_{i}\in\{\mathbbm{1},\gamma_5,\gamma^{\mu},\gamma_5\gamma^{\mu}, \sigma^{\mu\nu},g^{\mu\nu},\epsilon^{\mu\nu\rho\sigma}\}$ to keep the complete Clifford Algebra. While in Ref.~\cite{Petschauer:2013uua}, $\Gamma_{i}^{\,\prime}\in\{\mathbbm{1},\gamma_5\gamma^{\mu}, \sigma^{\mu\nu},g^{\mu\nu},\epsilon^{\mu\nu\rho\sigma}\}$ by means of the EOM. Nevertheless, these are simply different choices and can be transformed into each other using the EOM. Note that  $-2i{ m }\left(\bar{\psi}\gamma_5 \psi\right) \approx \partial^{\mu}\left(\bar{\psi} \gamma_5 \gamma_{\mu}\psi\right)$ and $2{m}\left(\bar{\psi}\gamma^{\mu} \psi\right)\approx \left(\bar{\psi} i\overleftrightarrow{\partial}^{\mu}\psi\right)$. The two structures on the left-hand-side of the above equations are contained in our work while the structures on the right-hand-side are included in Ref.~\cite{Petschauer:2013uua}.

  As in the one-baryon sector, in the two-nucleon sector, one also encounters the so-called power-counting-breaking problem~\cite{Gasser:1987rb}. Namely,  nominally higher order terms contain lower order terms that break
 the power counting. For instance, according to our criteria listed above, we should include $\widetilde{O}_{14}^{'}=\frac{1}{4m^2}\left( \bar{\psi} i\overleftrightarrow{\partial}^{\mu} \psi \right)\left( \bar{\psi} i\overleftrightarrow{\partial}_{\mu} \psi \right)$ at order $\mathcal{O}(q^2)$. However, this term actually starts to contribute at  $\mathcal{O}(q^0)$ so that it breaks our power counting. Therefore, we redefine $\widetilde{O}_{14}=\frac{1}{4m^2}\left( \bar{\psi} i\overleftrightarrow{\partial}^{\mu} \psi \right)\left( \bar{\psi} i\overleftrightarrow{\partial}_{\mu} \psi \right)-\left( \bar{\psi}  \psi \right)\left( \bar{\psi} \psi \right)$ to recover the power counting. The same also applies to $\widetilde{O}_{14-17}$  and $\widetilde{O}_{28-40}$. Notice that
 this procedure is very similar to the EOMS scheme~\cite{Gegelia:1999gf,Fuchs:2003qc},~\footnote{For a related discussion
 in the $NN$ sector, see, e.g., Ref.~\cite{Djukanovic:2007zz}.} except in the one baryon sector, power counting breaking terms only appear in the loop calculation with propagating baryons. In
 the nucleon-nucleon sector, they already appear at tree level because  the nucleon momentum is involved in order to increase the chiral order.

\subsection{Non-relativistic reductions}
The relativistic results, when reduced to the non-relativistic ones in the heavy baryon limit, should recover the well-known $2+7+15$ linear independent non-relativistic terms up to $\mathcal{O}(q^4)$~\cite{Epelbaum:2004fk}. We checked
that this is  indeed the case.

To perform the non-relativistic reduction of the covariant chiral Lagrangian constructed in our present work, one has to replace the relativistic nucleon field operator $\psi$ with the non-relativistic nucleon field operator $N$ and then expand the relativistic Lagrangian in terms of $1/m$.
The relativistic nucleon field operator $\psi(x)$ is~\footnote{Following Ref.~\cite{Girlanda:2010ya}, its negative-energy component has been dropped for simplicity.}
\begin{equation}
  \psi(x)= \int \frac{d {\bm p}}{(2 \pi)^3}
   \frac{m}{E_p}\, \widetilde{b}_{s}({\bm p}) \,u^{(s)}({\bm p})\,
  {\mathrm{e}}^{-i p\cdot  x} ,
\end{equation}
with the following normalization:
\begin{equation}
\left[ \widetilde{b}_s({\bm p}), \widetilde{b}^{\dagger}_{s^{\,\prime}}(\bm p^{\,\prime}) \right]_{+}=\frac{E_p}{m} (2\pi)^3 \delta({\bm p}-{\bm p^{\,\prime}}) \delta_{s s^{\,\prime}},~~ \bar{u}^{(s)}({\bm p}) u^{(s^{\,\prime})}({\bm p})=\delta_{ss^{\,\prime}},
\end{equation}
where $\widetilde{b}_s({\bm p})$ and  $\widetilde{b}^{\dagger}_{s^{\,\prime}}$ are annihilation and creation operators for a nucleon in spin state $s$ and $s^{\,\prime}$. A sum over the repeated index $s(s^{\,\prime})=\pm \frac{1}{2}$ is implied.
The Dirac spinors of $u$ and $\bar{u}$
have  the following form
\begin{equation}
 \bar u_i (\bm p',s) = \sqrt{\frac{E_i+m}{2m}}\left( \mathbbm1 \ ,\ -\frac{{\bm\sigma}\cdot{\bm p}^{\,\prime}}{E_i+m} \right)\chi_s^\dagger\,,\quad
 u_j (\bm p, s')= \sqrt{\frac{E_j+m}{2m}}\begin{pmatrix} \mathbbm1 \\ \frac{{\bm\sigma}\cdot{\bm p}}{E_j+m} \end{pmatrix}\chi_{s'} \,,
\end{equation}
with
\begin{equation}
 E_i = \sqrt{m^2 + {\bm p}^{\,\prime2}}\,,\quad
 E_j = \sqrt{m^2 + {\bm p}^{\,2}}\,.
\end{equation}
The non-relativistic nucleon filed $N(x)$ is
\begin{equation}
N(x) = \int \frac{d{\bm p}}{(2\pi)^3}\, b_s({\bm  p}) \, \chi_s\, {\rm e}^{-i p \cdot x} \ .
\label{eq:nrf}
\end{equation}
Here, the isospin indices have been suppressed by using the Fierz rearrangement~\cite{Girlanda:2010ya}. Note that $b_{s}({\bm p})=\sqrt{m / E_p} ~ \widetilde{b}_{s}({\bm p})$. To order $\mathcal{O}(q^4)$ at which we are working, the relativistic nucleon field operator $\psi$ can be expanded in terms of the non-relativistic field $N(x)$, defined in Eq.~(\ref{eq:nrf}), as
\begin{equation}
  \psi(x) =\left[ \left( \begin{array}{c} 1 \\
                                           0 \end{array} \right) -
   \frac{i}{2m}\left( \begin{array}{c} 0 \\
  {\bm \sigma} \cdot {\bm \nabla} \end{array} \right)
  +\frac{1}{8m^2}
   \left( \begin{array}{c} {\bm \nabla}^2\\
                              0 \end{array} \right)\right.
-\left.  \frac{3i}{16m^3}\left( \begin{array}{c} 0 \\
  {\bm \sigma} \cdot {\bm \nabla} {\bm \nabla}^2\end{array} \right)
  +\frac{11}{128m^4}
   \left( \begin{array}{c} {\bm \nabla}^4\\
                              0 \end{array} \right)\right] N(x) + {\mathcal O}(q^5)  \ .\label{eq:psiN}
\end{equation}
The Dirac matrices are defined as,
\begin{equation}
  \gamma^0 =  \left( \begin{array}{cc} 1 &0 \\
                                           0 &-1 \end{array} \right),
                                           \gamma^5= \left( \begin{array}{cc} 0 &1 \\
 1 &0 \end{array} \right),
 {\bm \gamma}=
   \left( \begin{array}{cc} 0 &{\bm{\sigma}}\\
                            -{\bm{\sigma}} & 0 \end{array} \right), \sigma^{\mu\nu}=\frac{i}{2}\left[\gamma^\mu,\gamma^\nu\right]   \label{eq:gammamatrix}
\end{equation}
with the Pauli matrices
\begin{eqnarray}
  &&\sigma_1 =  \left( \begin{array}{cc} 0 &1 \\
                                           1 &0 \end{array} \right),
                                           \sigma_2= \left( \begin{array}{cc} 0 &-i \\
 i &0 \end{array} \right),
 \sigma_3=
   \left( \begin{array}{cc} 1 &0\\
                            0 & -1 \end{array} \right).\\\nonumber
\label{eq:paulimatrix}
\end{eqnarray}
Using these relations and the properties of Pauli matrices, one can perform the non-relativistic expansion of  the covariant Lagrangian of Table~\ref{tb:NNLO NN Lagrangian}. The results are  summarized in Table~\ref{tb:NNLO NR RD}. They are presented
 as linear combinations of the non-relativistic Lagrangian $O_{i}$  defined in Table~\ref{tb:NNLO NR}. One can easily check  that there are $24$ linear independent terms, consistent with the well-known $2+7+15$ non-relativistic terms shown in Table~\ref{tb:NNLO NR}.

\begin{table}
  \caption{The non-relativistic expressions corresponding to the contact interactions of Table~\ref{tb:NNLO NN Lagrangian}.}
 \label{tb:NNLO NR RD}
 \centering
\rotatebox{90}{
  \begin{tabular}{c |c}
   \hline
   \hline
   $\widetilde{{O}}_1$    & $O_S+\frac{1}{4m^2}(O_1-2 O_2+2 O_3)+\frac{1}{16m^4}(3 O_8-4 O_9+O_{10}+O_{11}+4 O_{12}-2 O_{13}-O_{16}+O_{17}-O_{21}-O_{22})$ \\
   $\widetilde{{O}}_2$    &  $O_S+\frac{1}{4m^2}(4 O_2-6 O_3+O_4+2 O_5+O_6+2 O_7)+\frac{1}{16m^4}(8 O_9-O_{10}+O_{11}-12 O_{12}-2 O_{13}+3 O_{14}+4 O_{15}+O_{16}+O_{17}+3 O_{18}+4 O_{19}-O_{22})$\\
   $\widetilde{{O}}_3$    & $-O_T+\frac{1}{4m^2}(-2 O_3-O_4+2 O_5-O_6+6 O_7)+\frac{1}{16m^4}(O_{10}-O_{11}-4 O_{12}+2 O_{13}-3 O_{14}+4 O_{15}-O_{16}-O_{17}-3 O_{18}+12 O_{19}-2 O_{22})$ \\
   $\widetilde{{O}}_4$    & $2 O_T+\frac{1}{4m^2}(2 O_1+4 O_2-12 O_3+8 O_5-2 O_6+12 O_7)+\frac{1}{16m^4}(6 O_8+8 O_9+2 O_{10}+2 O_{11}-24 O_{12}-4 O_{13}+16 O_{15}-2 O_{16}+2 O_{17}-6 O_{18}+24 O_{19}-4 O_{21}+4 O_{22})$  \\
   $\widetilde{{O}}_5$    &  $-\frac{1}{4m^2}(O_6+2 O_7)-\frac{1}{16m^4}(3 O_{18}+4 O_{19}+O_{21})$\\
   $\widetilde{{O}}_6$    &  $\frac{1}{4m^2}(-4 O_6+8 O_7)+\frac{1}{16m^4}(-4 O_{18}+4 O_{21})$\\
   $\widetilde{{O}}_7$    &  $\frac{1}{4m^2}(O_1+2 O_2-8 O_3-4 O_4+8 O_5-4 O_6+8 O_7)+\frac{1}{16m^4}(-4 O_{10}+4 O_{11}+2 O_{12}-12 O_{13}-4 O_{14}+8 O_{17}-4 O_{18}-4 O_{21}-8 O_{22})$\\
   $\widetilde{{O}}_{8}$  & $\frac{1}{4m^2}(O_1+2 O_2-4 O_3)+\frac{1}{16m^4}(-2 O_{12}+4 O_{13}+4 O_{16}-4 O_{17}+4 O_{21}+4 O_{22})$ \\
   $\widetilde{{O}}_{9}$  &  $\frac{1}{4m^2}\left(-O_4-2 O_5-O_6-2 O_7\right)+\frac{1}{16m^4}\left(-O_8-4 O_9-2 O_{10}-4 O_{11}+6 O_{12}+12 O_{13}-O_{14}-4 O_{15}+2 O_{16}-8 O_{17}-O_{18}-4 O_{19}+5 O_{21}+8 O_{22}\right)$ \\
   $\widetilde{{O}}_{10}$    & $\frac{1}{4m^2}(-O_1-2 O_2)+\frac{1}{16m^4}(-2 O_8+4 O_{11}-2 O_{12}-4 O_{13})$ \\
   $\widetilde{{O}}_{11}$    & $\frac{1}{4m^2}\left(-O_1-2 O_2\right) + \frac{1}{16m^4}\left(-O_8-4 O_9+2 O_{10}-8 O_{11}+6 O_{12}+12 O_{13}-O_{14}-4 O_{15}-2 O_{16}-4 O_{17}-O_{18}-2 O_{19}-2 O_{20}-O_{21}-4 O_{22} \right)$ \\
   $\widetilde{{O}}_{12}$    &  $\frac{1}{4m^2}(O_4+2 O_5)+\frac{1}{16m^4}(2 O_{12}+4 O_{13}+2 O_{14}-4 O_{17}+O_{18}-6 O_{19}+2 O_{20}+O_{21}-12 O_{22})$\\
   $\widetilde{{O}}_{13}$    & $\frac{1}{4m^2}(-2 O_4-4 O_5)+\frac{1}{16m^4}(-2 O_8-8 O_9-4 O_{10}-8 O_{11}+12 O_{12}+24 O_{13}-2 O_{14}-8 O_{15}+4 O_{16}-16 O_{17}+2 O_{18}-12 O_{19}+4 O_{20}+2 O_{21}-24 O_{22})$ \\
   $\widetilde{{O}}_{14}$   & $\frac{1}{4m^2}(-3 O_1+2 O_2)+\frac{1}{16m^4}(-4 O_8+8 O_9-4 O_{10}-4 O_{11}-6 O_{12}+4 O_{13})$ \\
   $\widetilde{{O}}_{15}$   & $\frac{1}{4m^2}(-3 O_1+2 O_2)+\frac{1}{16m^4}(-O_8-12 O_9+2 O_{10}+8 O_{11}+18 O_{12}-12 O_{13}-3 O_{14}-4 O_{15}-6 O_{16}+4 O_{17}-3 O_{18}-6 O_{19}+2 O_{20}-3 O_{21}+4 O_{22})$ \\
   $\widetilde{{O}}_{16}$   & $\frac{1}{4m^2}(3 O_4-2 O_5)+\frac{1}{16m^4}(6 O_{12}-4 O_{13}+4 O_{14}-8 O_{15}+4 O_{16}+4 O_{17}+3 O_{18}-18 O_{19}-2 O_{20}+3 O_{21}+12 O_{22})$ \\
   $\widetilde{{O}}_{17}$   & $\frac{1}{4m^2}(-6 O_4+4 O_5)+\frac{1}{16m^4}(-6 O_8-8 O_9-12 O_{10}+8 O_{11}+36 O_{12}-24 O_{13}-2 O_{14}-24 O_{15}+4 O_{16}+16 O_{17}+6 O_{18}-36 O_{19}-4 O_{20}+6 O_{21}+24 O_{22})$ \\
   $\widetilde{{O}}_{18}$   & $\frac{1}{16m^4}(O_{18}+2 O_{19}+2 O_{20}+O_{21}+4 O_{22})$ \\
   $\widetilde{{O}}_{19}$   & $\frac{1}{16m^4}(4 O_{18}-8 O_{19}+8 O_{20}+4 O_{21}-16 O_{22})$ \\
   $\widetilde{{O}}_{20}$   & $\frac{1}{16m^4}(O_8+4 O_9+2 O_{10}+4 O_{11}-8 O_{12}-16 O_{13}-4 O_{14}-8 O_{16}+16 O_{17}-4 O_{18}+8 O_{19}-8 O_{20}-4 O_{21}+16 O_{22})$ \\
   $\widetilde{{O}}_{21}$   & $\frac{1}{16m^4}(-O_8-4 O_9-2 O_{10}-4 O_{11}+4 O_{12}+8 O_{13})$ \\
   $\widetilde{{O}}_{22}$   & $\frac{1}{16m^4}(O_{14}+4 O_{15}+2 O_{16}+4 O_{17}+O_{18}+2 O_{19}+2 O_{20}+O_{21}+4 O_{22})$ \\
   $\widetilde{{O}}_{23}$   & $\frac{1}{16m^4}(-O_8-4 O_9-2 O_{10}-4 O_{11}+8 O_{12}+16 O_{13}+16 O_{16}-16 O_{17}+16 O_{21}+16 O_{22})$ \\
   $\widetilde{{O}}_{24}$   & $\frac{1}{16m^4}(O_8+4 O_9+2 O_{10}+4 O_{11})$ \\
   $\widetilde{{O}}_{25}$   & $\frac{1}{16m^4}(O_8+4 O_9+2 O_{10}+4 O_{11})$ \\
   $\widetilde{{O}}_{26}$   & $\frac{1}{16m^4}(-O_{14}-4 O_{15}-2 O_{16}-4 O_{17})$ \\
   $\widetilde{{O}}_{27}$   & $\frac{1}{16m^4}(2 O_{14}+8 O_{15}+4 O_{16}+8 O_{17})$ \\
   $\widetilde{{O}}_{28}$   & $\frac{1}{16m^4}(3 O_{18}+6 O_{19}-2 O_{20}+3 O_{21}-4 O_{22})$ \\
   $\widetilde{{O}}_{29}$   & $\frac{1}{16m^4}(12 O_{18}-24 O_{19}-8 O_{20}+12 O_{21}+16 O_{22})$ \\
   $\widetilde{{O}}_{30}$   & $\frac{1}{16m^4}(-3 O_8-4 O_9-6 O_{10}+4 O_{11}+24 O_{12}-16 O_{13}+12 O_{14}-32 O_{15}+24 O_{16}+16 O_{17}+12 O_{18}-24 O_{19}-8 O_{20}+12 O_{21}+16 O_{22})$ \\
   $\widetilde{{O}}_{31}$   & $\frac{1}{16m^4}(-O_8-4 O_9-2 O_{10}-4 O_{11}+4 O_{12}+8 O_{13})$ \\
   $\widetilde{{O}}_{32}$   & $\frac{1}{16m^4}(3 O_{14}+4 O_{15}+6 O_{16}-4 O_{17}+3 O_{18}+6 O_{19}-2 O_{20}+3 O_{21}-4 O_{22})$ \\
   $\widetilde{{O}}_{33}$   & $\frac{1}{16m^4}(3 O_8+4 O_9+6 O_{10}-4 O_{11})$ \\
   $\widetilde{{O}}_{34}$   & $\frac{1}{16m^4}(3 O_8+4 O_9+6 O_{10}-4 O_{11})$ \\
   $\widetilde{{O}}_{35}$   & $\frac{1}{16m^4}(-3 O_{14}-4 O_{15}-6 O_{16}+4 O_{17})$ \\
   $\widetilde{{O}}_{36}$   & $\frac{1}{16m^4}(6 O_{14}+8 O_{15}+12 O_{16}-8 O_{17})$ \\
   $\widetilde{{O}}_{37}$   & $\frac{1}{16m^4}(9 O_8-12 O_9+18 O_{10}+4 O_{11})$ \\
   $\widetilde{{O}}_{38}$   & $\frac{1}{16m^4}(9 O_8-12 O_9+18 O_{10}+4 O_{11})$ \\
   $\widetilde{{O}}_{39}$   & $\frac{1}{16m^4}(-9 O_{14}+12 O_{15}-18 O_{16}-4 O_{17})$ \\
   $\widetilde{{O}}_{40}$   & $\frac{1}{16m^4}(18 O_{14}-24 O_{15}+36 O_{16}+8 O_{17})$ \\
   \hline
   \hline
  \end{tabular}}
\end{table}

\section{ Summary}
We have constructed a complete set of  relativistic $NN$ contact Lagrangian terms up to order $\mathcal{O}(q^4)$. Using the EOM to eliminate the redundant terms, we find only 4 terms of $\mathcal{O}(q^0)$, 13 terms of $\mathcal{O}(q^2)$,
 and {23} terms of $\mathcal{O}(q^4)$. We compared with previous studies and identified the differences and the reasoning behind them.  In addition, we checked that
 by performing $1/m_N$ expansions, one can recover the corresponding non-relativistic chiral Lagrangians, which has 24 terms up to $\mathcal{O}(q^4)$. The covariant Lagrangian constructed in
 the present work can be of use in building covariant nuclear forces as well as studying relativistic corrections.

 It should be emphasized that the completeness and minimality of the set of covariant Lagrangian terms derived in the present work should be understood
  with respect to the power counting rules we chose and the choice we made regarding how to use
 the equation of motion to remove redundant terms and to limit the number of available terms. More precisely, they may better be referred to  as an economical set of covariant Lagrangian terms that can recover those of the heavy baryon.
 We hope that the present work can motivate more studies in this direction.

\section{Acknowledgement}
YX and LSG thank Dr. Shao-Zhou Jiang for a careful reading of the manuscript and for many useful comments.  YX thanks Dr. Stefan Petschauer for useful discussions and acknowledges the hospitality of Technique University of Munich where part of this work is done.
 This work is partly supported by the National Natural Science Foundation of China under Grants No.11522539, No. 11735003, No. 11775099, and by the fundamental Research Funds for the Central Universities. XLR acknowledges supports from DFG and NSFC through funds provided to the Sino-German CRC 110 ``Symmetries and the Emergence of Structure in QCD'' (Grant No. TRR110).

%\appendix{Non-relativistic $NN$ contact Lagrangian up to order $\mathcal{O}(q^4)$}
\section{Appendix}

In this section, we construct the non-relativistic Lagrangian up to order $\mathcal{O}(q^4)$. Compared to the covariant case, the procedures to construct the non-relativistic chiral Lagrangian is simple and straightforward. The basic requirement for
a non-relativistic Lagrangian is that the Lagrangian must be scalar.  Because of parity constraints, the non-relativistic Lagrangian must contain an even number of gradient operators. Because the three momentum of the nucleon is the only small expansion parameter in the
 non-relativistic power counting~\footnote{We do not consider external fields in this work.}, to the $n$th order, one needs to include $n$~gradient operators. Using the criteria listed above,  one can easily construct the non-relativistic Lagrangians up to order $\mathcal{O}(q^4)$. They  are summarized in Table~\ref{tb:NNLO NR},
 where $N$ and $N^\dagger$ donate the nucleon field and its hermitian conjuage, and $ {\bm\nabla}$ refers to the gradient operator. Note that $(N^\dagger {\bm \nabla} N)(N^\dagger N)=-(N^\dagger N)(N^\dagger {\bm \nabla} N)$ since we are working in the center-of-mass frame.

\begin{table}[h]
\caption{ $\mathcal{O}(q^0)$, $\mathcal{O}(q^2)$ and $\mathcal{O}(q^4)$ non-relativistic $NN$ contact Lagrangian terms.  The left (right) arrow on ${\bm\nabla}$  indicates that the derivative  acts on the left (right) nucleon field.}
\label{tb:NNLO NR}
\centering
\begin{tabular}{c|c|c|c}
\hline
\hline
$O_S$  & $(N^\dagger N)(N^\dagger N)$& $O_{11}$ & $(N^\dagger  \overrightarrow{\bm \nabla} \cdot \overleftarrow{\bm \nabla} N)(N^\dagger \overrightarrow{\bm \nabla} \cdot \overleftarrow{\bm \nabla} N)$\\
$O_T$  & $(N^\dagger {\bm {\sigma}} N)\cdot(N^\dagger {\bm {\sigma}} N)$& $O_{12}$ & $i\, (N^\dagger {\bm \sigma} \cdot \overrightarrow{\bm \nabla}\times \overleftarrow{\bm \nabla} N)(N^\dagger \overrightarrow{\bm \nabla}^2  N)+{\rm h.c.}$\\
\cline{1-2}
$O_1$  &  $(N^\dagger N) ( N^\dagger \overrightarrow{ \bm\nabla}^2 N)+{\rm h.c.}$    &$O_{13}$ & $i\, (N^\dagger {\bm \sigma} \cdot \overrightarrow{\bm \nabla}\times \overleftarrow{\bm \nabla} N)(N^\dagger \overrightarrow{\bm \nabla} \cdot \overleftarrow{\bm \nabla}  N)$\\
$O_2$  & $(N^\dagger  N )( N^\dagger  \overrightarrow{\bm \nabla} \cdot \overleftarrow{\bm \nabla} N)$ &$O_{14}$ & $ (N^\dagger \sigma^{j} N)(N^\dagger  \sigma^{j} \overrightarrow{\bm \nabla}^4 N)+{\rm h.c.}$\\
$O_3$  & $i \,( N^\dagger {\bm{\sigma}} N) \cdot (N^\dagger \overrightarrow{\bm \nabla}
  \times \overleftarrow {\bm \nabla} N )$ &$O_{15}$ & $ (N^\dagger \sigma^{j} \overrightarrow{\bm \nabla} \cdot \overleftarrow{\bm \nabla}N)(N^\dagger  \sigma^{j} \overrightarrow{\bm \nabla}^2 N)+{\rm h.c.}$ \\
$O_4$ & $ (N^\dagger {\sigma}^{j} N)(N^\dagger {\sigma}^{j} \overrightarrow{\bm \nabla}^2 N)+{\rm h.c.}$ &$O_{16}$ & $ (N^\dagger \sigma^{j} \overrightarrow{\bm \nabla}^2 N)(N^\dagger  \sigma^{j} \overleftarrow{\bm \nabla}^2 N)$\\
$O_5$ & $ (N^\dagger {\sigma}^{j} N)(N^\dagger {\sigma}^{j} \overrightarrow{\bm \nabla}\cdot \overleftarrow{\bm \nabla} N)$ &$O_{17}$ & $ (N^\dagger \sigma^{j} \overrightarrow{\bm \nabla} \cdot \overleftarrow{\bm \nabla} N)(N^\dagger  \sigma^{j} \overrightarrow{\bm \nabla} \cdot \overleftarrow{\bm \nabla}  N)$\\
$O_6$ & $ (N^\dagger {\bm{ \sigma}}\cdot \overrightarrow{\bm \nabla} N)  (N^\dagger {\bm{ \sigma}}\cdot \overrightarrow{\bm \nabla} N)+{\rm h.c.}$ &$O_{18}$ & $ (N^\dagger {\bm\sigma} \cdot \overrightarrow{\bm \nabla}  N)(N^\dagger  {\bm \sigma} \cdot \overrightarrow{\bm \nabla}  \overrightarrow{\bm \nabla}^2  N)+{\rm h.c.}$ \\
$O_7$ & $ (N^\dagger {\bm{ \sigma}}\cdot \overrightarrow{\bm \nabla} N)  (N^\dagger {\bm{ \sigma}}\cdot \overleftarrow{\bm \nabla} N)$ &$O_{19}$ & $ (N^\dagger {\bm\sigma} \cdot \overrightarrow{\bm \nabla}  N)(N^\dagger  {\bm \sigma} \cdot \overleftarrow{\bm \nabla}  \overrightarrow{\bm \nabla}^2  N)+{\rm h.c.}$\\
\cline{1-2}
$O_8$ & $(N^\dagger N)(N^\dagger \overrightarrow{\bm \nabla}^4 N)+{\rm h.c.}$ & $O_{20}$ & $ (N^\dagger {\bm\sigma} \cdot \overrightarrow{\bm \nabla}  N)(N^\dagger  {\bm \sigma} \cdot \overrightarrow{\bm \nabla}  \overrightarrow{\bm \nabla} \cdot \overleftarrow{\bm \nabla}  N)+{\rm h.c.}$\\
$O_9$ & $(N^\dagger \overrightarrow{\bm \nabla}^2 N)(N^\dagger  \overrightarrow{\bm \nabla} \cdot \overleftarrow{\bm \nabla} N)+{\rm h.c.}$ &$O_{21}$ & $ (N^\dagger {\bm\sigma} \cdot \overrightarrow{\bm \nabla}  N)(N^\dagger  {\bm \sigma} \cdot \overrightarrow{\bm \nabla}  \overleftarrow{\bm \nabla}^2  N)+{\rm h.c.}$\\
$O_{10}$ & $(N^\dagger  \overrightarrow{\bm \nabla}^2 N)(N^\dagger \overleftarrow{\bm \nabla}^2 N)$ &$O_{22}$ & $ (N^\dagger {\bm\sigma} \cdot \overrightarrow{\bm \nabla}  N)(N^\dagger  {\bm \sigma} \cdot \overleftarrow{\bm \nabla}  \overrightarrow{\bm \nabla} \cdot \overleftarrow{\bm \nabla}  N)$\\
\hline
\hline
\end{tabular}
\end{table}

\bibliography{refs}
\end{document}